# Data Center Constructions in the United States: When and How does Local Resistance Succeed?

**Prof. Jason Miklian**
Centre for Global Sustainability, University of Oslo
jason.miklian@globe.uio.no

## Abstract

Residents across the United States are organizing against data centers proposed by some of the world's largest companies. Local governments have adopted 603 moratoria and other restrictions in 445 different counties through October 2026, most of which were enacted this year. The firms pitch the projects by offering jobs and tax revenue, while residents believe that they bear the brunt of the facilities' power demand, water use, noise, and societal consequences. However, only a portion of data centers have been stopped despite bipartisan citizen desire, reflected in the fact that all 42 data center proposals on ballots across the US in November 2026 were rejected. Therefore, when are local communities able to restrict data centers, what are the most successful strategies, and can we decipher the conditions for when this resistance succeeds? To answer, this paper inventories 2,191 data center projects across the United States, matching them to local policy records to establish when counties acted and which activities lead to construction, a moratorium, or outright cancellation. The analysis finds that successful resisters were most likely to adopt their first local restriction in the first six months after a project announcement. After one county adopted a restriction, restrictions were more frequent in neighboring counties as well, suggesting that local awareness has a strong proximity spillover effect. Projects in counties with no opposition action recorded in our sources had the largest construction gap after early restrictions, and also correlated with lower local media penetration, suggesting that access to local media helps increase awareness of projects, triggering more successful resistance via moratoria and engagement after projects are permitted. A firm's threat to build elsewhere gives it bargaining power over officials who want its investment, but since residents do not receive the typical benefits from such sorts of project (e.g. jobs), the power of this forum shopping threat is diminished. Taken together, these findings show where and why new data center projects succeed or are shelved, building new evidence for scholars, policymakers, and the public on this polarizing issue.

## 1. Introduction

The promise of data center investment leaves local governments and communities with a consequential choice: how much of their own resources should their communities commit to help these actors thrive? Birmingham, Alabama is in many ways a typical case. Upon hearing that a large data center was planned in their community, residents and environmental advocates pressed officials to pause new projects, raising concerns about how the projects promised little return in exchange for large-scale facilities that made heavy utility demands and noise. Following a public hearing on March 3, 2026, the council unanimously approved a 180-day pause on new hyperscale data center applications, allowing time to establish those conditions with public input (Bowser 2026; City of Birmingham 2026). Birmingham's decision reflects a growing national movement to halt or reconsider the benefits of data center expansion. By October 2026, local governments in 445 different American counties had adopted 603 restrictions on data centers since January 2017, including 534 in 2026 alone.

In the usual account of local development politics, officials and the people who own and develop land want economic and community growth. They thus compete for projects that bring private investment, jobs, and taxable property into their jurisdiction (Molotch 1976; Peterson 1981). Research on business power ties corporate influence to low public attention (Culpepper 2010), and developers describe pressure to implement features that conflict with their own social values (Miklian and Hoelscher 2025). As data centers attracted public opposition, councils adopted hundreds of restrictions, sometimes even in counties with no forthcoming project at all (Stokes et al. 2023; Reece et al. 2026). This wide scale resistance surprised technology firms who had grown accustomed to using traditional forum shopping to their benefit (typically by threatening to build) elsewhere, unable to align their projects convincingly with what the people deciding on their permits want or need in return.

In the standard account, a threat to move gives the firm power because local officials want to retain the investment and the revenue it brings (Peterson 1981). Residents and council members seeking to keep a facility out have the opposite objective: forbid their approval until the firm makes concessions that benefit the community (or at least the vocal parts of it). Before permits are issued, officials decide if and how to approve this development; later, they must also deal with how granted approvals have been applied. Land purchases and contracts then commit the developer to the site, and permits establish rights under the rules in force when they are issued. Some of these commitments are made before the public even hears about a project, through confidential site selection and negotiations over infrastructure and incentives (Wood and Phelps 2020; Kollar 2026a). Officials seeking such investment then may compete to retain it by offering tax incentives, a common part of negotiations over location (Jensen 2018).

Large data centers are a new type of large-scale development that stress this understanding. Their locations are much more fungible than most other projects, but their local gains through jobs and taxes are typically slim, while at the same time they can carry significant negative local environmental and economic effects. The neighbors of a data center must contend with its power demand, water use, and noise, giving them reasons to oppose a project their officials want to improve tax revenues and project their communities as future-oriented and thus worthy of additional investment by technology actors.

Recently, data centers have become lightning rods for advocates against both unplanned local development and the role of Artificial Intelligence (AI) companies in society. Given that the companies that build these data centers are some of the world's largest firms, it can look like a David vs. Goliath battle that a local community would be unlikely to succeed in. Yet, a huge number of restrictions have indeed been passed all around the United States, and hundreds of projects have indeed been stopped or relocated. This leads to two questions: When are local governments able to restrict data centers, and do the successful restriction events carry certain commonalities that can be ascertained?

These questions inform our main hypotheses. Given the groundswell against AI development across the country, we expect counties to consider restrictions soon after a new data project is announced, when the proposal first attracts public attention and officials face their first decisions about its development (H1). Moreover, local governments preparing for possible future investment also have reasons to consider the adoption of rules (in this case moratoria) even if a new project is not imminent (H2). We propose that restrictions should spread across adjacent or nearby counties where the same opposition group operates, since those connections give officials and residents more access to experience from other disputes than media or social media might expose them to (H3). To test these hypotheses, we assembled an inventory of 2,191 data center projects from four public sources and matched these to the 603 restrictions across all 3,222 U.S. counties and county equivalents. Then, we compared outcomes for projects in counties that adopted early restrictions and for other projects announced in the same period, from initial notification to permitting through to construction.

The analysis found that local governments were most likely to adopt data center restrictions in the first six months after a project is announced, and restrictions adopted in those first six months are most likely to be effective in halting eventual construction compared to restrictions attempted after six months. Also, counties that are next to or share organizations with counties that had restrictions were more likely to successfully restrict projects themselves. These policies, where enacted, also shaped the conditions for future development, as the largest construction gap between planning and groundbreaking appears where governments adopt early restrictions without previously recorded opposition. This places council decisions and permitting powers alongside public protest as central forms of local early intervention, in effect planting an early flag for AI companies to read that this location will be harder than others to succeed. This proactivity constitutes a new wrinkle on the forum shopping literature, showing how local communities can preemptively take themselves out of the running as a possible forum altogether. Last, the media and timing evidence impacts what we understand to be conflict timelines. In "news deserts", local communities in many cases did not hear about proposed developments within the crucial 6-month window, reducing their likelihood to successfully resist. But in 43 counties, local news covered data center disputes before the recorded announcement and success rates were higher than average (albeit without enough data points to be considered significant), speaking to the importance of the existence of local media sources to raise knowledge about industry actions.

## 2. Theory

### *2.1 Mobile capital, local refusal and timing*

In many traditional arguments about county and municipal capital mobility, it is assumed that capital and labor move across local borders, giving local governments a reason to pursue development because their tax base depends on it (Peterson 1981). The growth machine literature describes the coalition behind that pursuit as landowners, developers, and allied officials who share an interest in more intensive land use for economic exploitation (Molotch 1976; Logan and Molotch 2007). Landowners gain from higher property values and developers from new building, while the costs of development fall on a wider population. The competition for investment can also give firms access to public subsidies that may not otherwise be economically beneficial. State and local governments have often paid incentives for location decisions that firms had already made (Jensen 2018; Slattery and Zidar 2020). Officials keep offering them partly because attracting a corporate headquarters affects how voters judge incumbents (Yang 2024), and partly because of the prestige factor that certain types of firms provide for the region and in turn of its elected officials (such as professional sports teams).

The standard account assumes (with some exceptions, e.g. a waste dump) that the people deciding on a project want the investment in principle, but the terms and benefits of the project are the point of negotiation. These actors choose to allow or withhold permission from a proposed facility. Who controls that choice varies across places: officials weigh general welfare against benefits concentrated among influential constituencies, and dependence on an established industry shapes both government capacity and the firm's bargaining position (Bhaumik et al. 2024; Elder 2025). Securing an agreement thus requires coordination with local government over planning and infrastructure (Ramos Cáceres et al. 2024).

Residents who organize against a project often want divergent and even contradictory things. Some want it kept out entirely, some want time to examine the proposal, and others want better terms. The consequences are thus also different: a pause answers a demand for time while abandoning the development answers a demand for exclusion. Social movement research is a rich body of literature that compares outcomes with participants' demands and considers what agreeing to those demands would cost (Amenta et al. 2005; Luders 2010), as in research on housing and wind farms that studies who is allowed to participate in decisions (Einstein et al. 2019; Stokes et al. 2023). Table 1 distinguishes these outcomes accordingly:

*Table 1. Typical outcomes to local project discussions and evidence to verify such*

| Outcome | Local Action | Verifying records |
|---|---|---|
| **Regulatory adoption** | A local government adopts a restriction | A dated ordinance or other policy instrument |
| **Binding project coverage** | The restriction applies to a particular project | Authority, project location, regulated use, effective dates, exemptions, application history |
| **Delay** | A project takes longer to reach a development milestone | Dated milestones and a comparison of development timelines |
| **Permanent prevention** | The proposed development is abandoned | Project history confirmed through extended follow-up |
| **Conditional approval** | A project is approved under changed terms | Documents specifying the revised terms or obligations |

However, a deliberative body deciding whether to admit a new data center facility may be considered to be facing an earlier stage of that relationship. Corporate spending tied to a site begins with options on land and agreements over infrastructure, giving developers reasons to defend the location before construction starts. For a data center, officials who support the project have reasons to protect those decisions and move quickly before resistance coalesces. To wit, Google’s proposed data center in Luxembourg stayed suspended for years while the firm retained its options and the unfinished development became an issue in local politics (Carr and Madron 2025). Much of this preparation occurs out of public view through confidential negotiations and site-selection intermediaries, and disputes over disclosure also surround incentive renegotiations (Wood and Phelps 2020; Jensen and Thrall 2021). This non-public engagement has been a tool that AI companies have used to their advantage. In Oregon, fiscal incentives, utility arrangements and land-use decisions were all negotiated as parts of the same development process, extending beyond the planning hearing to help make the project present more like a *fait accompli* (Kollar 2026a).

Residents learn about a proposal at some point in this process, often but not always after negotiations have begun. This distinction matters as concerns what is deliberated upon. A council hearing objections before it grants permits is deciding whether to approve development at all; a council hearing them after such approvals must justify the approvals but also this may make the deliberations themselves about the terms of the approval as opposed to the approval itself. Public attention changes the conditions under which business exercises influence, and firms' own conduct can disrupt the political relationships on which that influence rests so it is in their interest to keep such negotiations non-public for as long as feasible (Culpepper 2010; Bell and Hindmoor 2024).

State governments can take the decision away from a council. State preemption, through which state law limits local authority, has expanded in response to local policy initiatives, and planning research identifies the same conflict over data center regulation (Riverstone-Newell 2017; Goodman et al. 2021; Goodman and Hatch 2024; Kollar 2026b). Within a jurisdiction, homeowners with time to attend hearings and a property interest in the outcome have disproportionate influence over land-use decisions (Einstein et al. 2019; Hankinson 2018; Marble and Nall 2021). A refusal also affects people beyond the jurisdiction when a developer seeks another site. Following the firm's subsequent proposals would show where that search leads and who faces the next development, extending the questions about unequal siting burdens examined in research on wind and polluting facilities (Stokes et al. 2023; Mohai and Saha 2015).

A firm's response to social demands also depends on whether those demands enter company decisions through binding rules or through political pressure (Miklian 2026b). Once a multinational has invested, its assets are tied to the location, shifting bargaining power toward the host government (Vernon 1971). There is a rich literature on disputes over where to build new projects, who opposes them, which groups succeed in delaying or stopping them, and under what conditions (Giordono et al. 2018; Dokshin 2016; Stokes et al. 2023). In New York, for example, residents were more likely to win a local fracking ban when they worried about nearby drilling and few landowners in town stood to earn money from gas leases (Dokshin 2016). We build upon these studies to interrogate the importance of when local governments adopt their first restriction, using the public announcement as the dated event within a longer process of spending, negotiation and approval. This leads to H1's specific timing test if a county's monthly rate of first restriction is higher in the six months beginning with a data center announcement than in later months, before announcement, or in counties with no recorded project.

*2.2 Preparation and the spread of restrictions*

Local governments also prepare and incorporate rules for development that they expect to (or desire to) host in the future. Research on data centers shows that pro-build local leaders can prepare land and institutions years before the anticipated investment arrives in preemptive signaling to companies (Burrell 2020). These dynamics cut across sectors. County wind ordinances evolved through experience with regulation, and research on solar development likewise examines where local ordinances govern construction (Winikoff 2022; Lee et al. 2026).

In disputes over proposed data centers, residents who speak at public meetings tend to be less supportive than local officials (Reece et al. 2026; Kollar 2026a) Preparing for future proposals gives officials elsewhere a reason to act as well, extending regulation beyond the sites listed in a project inventory. Therefore, H2 proposes that most counties that adopt a data center restriction have no imminent upcoming project recorded as these are also preparatory, not acute, questions.

Borrowing from other governments reduces the work of preparing a local policy. Officials are known to use another council or municipality's account of an issue and the ordinance it adopted, then decide how those rules fit their own jurisdiction (Berry and Berry 1990; Boushey 2010). Government actors also learn about a policy's political acceptability as well as its performance, developing a shared understanding of an issue (Gilardi et al. 2021). For a council preparing data center rules, another jurisdiction's moratorium supplies an example of both potential citizen sentiment as well as the practicalities of implementation. These dynamics can also happen at the state level, as some states repeatedly follow other states' lead on policy (Desmarais et al. 2015; Harden et al. 2023).

Opposition groups provide one route for officials and residents that want to resist to learn about restrictions elsewhere. One good corollary is in the fracking debate, where coalition partners and a group's standing determine influence and policy change (Arnold et al. 2017). Likewise, local restrictions on renewable energy cluster near earlier adopters, while policies also spread between states governed by the same party (Ko et al. 2023; DellaVigna and Kim 2026). A group working in two counties offers a connection through which residents and officials can exchange an ordinance or contact someone involved in the earlier dispute.

Are similar dynamics at play with data centers? We propose H3: after a county adopts a restriction, neighboring counties with similar opposition groups will have a higher rate of first restriction during the next three months. We also propose H4: we expect fewer construction starts where councils adopt restrictions to stop proposed facilities, with the largest difference where residents organized against development before the vote.

We therefore compare construction across counties with and without opposition, noting that the sequence from announcement to restriction involves several possible routes to a council vote. Residents press officials to act, officials prepare rules themselves, and financial or infrastructure problems affect projects while those decisions are under way. Reporting also brings projects, groups and policy decisions into the record together. We examine these sequences by dividing projects according to whether an opposition action was recorded in the county before its first restriction and by comparing restriction dates with cancellation dates. These comparisons establish which event was recorded first and identify cases for examining how officials reached their decision. Work on political mediation examines how movements change specific decisions, including whether officials respond to their demands (Amenta et al. 2005; King 2008). This informs H5: the greater the presence of local media, the more likely that citizens hear about proposed data centers in a timely fashion.

## 3. Data and Methods

### *3.1 Project and policy inventories*

We source facilities by address at the county level, and the policy sources identify the jurisdiction adopting an ordinance. The panel covers all 3,222 counties and county equivalents in the 50 states, the District of Columbia and Puerto Rico during the 118 months from January 2017 to October 2026. We assembled the list of proposed data center projects from Server Country, Compute Atlas, EPA air-permit and facility records, and Epoch AI (see supplement and replication package codebook). We then grouped entries referring to the same development into a single project record, which we call a cluster. The inventory contains 2,191 clusters, of which 2,156 are assigned to 703 counties (35 could not be assigned to a county). A cluster can include several phases of one campus. Where an operator or project name changed across sources, the release preserves the identifiers and source dates so readers can check how we linked the entries and classified continuation or cancellation (see supplement). Restrictions come from the Moratorium Nation inventory, which covers several industries (Bommarito 2026). Among the place-based records, 642 have a usable month within our study period, and 617 of those explicitly identify data centers as a regulated sector. Removing 14 pending records leaves 603 restrictions in 445 counties used in the main analysis; we restore those 14 in a second series of 617 records to check how they affect the results, comparing counties within the same calendar month to account for changes over time in both adoption and collection (SI S11).

To test H2, we classified the 445 adopting counties by their project histories at the month of first restriction. We include cancelled developments as they record a developer's interest in the county. A county is included in the prior-project group when at least one project was announced in or before the adoption month or was dated to an earlier year. The later-project group contains counties whose recorded projects all follow adoption. In a third group, project dates leave the sequence unresolved. The final group contains counties with no recorded project. Table 2 reports these four groups; see supplement for classification rules and results after restoring pending restrictions.

*3.2 Adoption panels and the construction comparison*

We also compare a county's monthly rate of first restriction before and after its first recorded project announcement. We began with 282 announcement counties and 2,519 comparison counties with no recorded project. We excluded 20 counties with undated restrictions, leaving 274 announcement counties and 2,507 comparison counties (see supplement). Each contributes one observation per month until its first restriction or October 2026. The outcome equals one in the month of first adoption and zero in earlier months, yielding 325 first adoptions in 326,374 county-months (see Table 3).

We estimate how the monthly adoption rate differs after a project announcement, clustering standard errors by county. The outcome is whether the county adopts its first restriction that month, and the coefficient expresses the difference in percentage points. A second regression divides the period after announcement into months zero to five, six to eleven, and twelve onward; Figure 1 plots the three estimates. Counties leave the sample after their first restriction, following event-history practice for studying first adoptions (Berry and Berry 1990). Supplement Table S4 repeats the comparison with state-by-month effects, county effects, state-clustered errors, and pending restrictions restored. We also date reports on data center disputes, assigning outlets to counties through the 3DLNews directory (Ariyarathne and Nwala 2024). For the 191 counties whose first recorded project appeared between January 2025 and August 2026, we separately searched for local or regional coverage identifying that project during its announcement month or the preceding six months.

The test of adoption among counties connected by opposition groups uses a larger sample: all 3,188 counties whose recorded restrictions have dates. It contains 373,804 county-months and 430 first adoptions, with each county again leaving the sample after its first restriction, identified through the Data Center Opposition Tracker. For each county-month, we count how many connected counties adopted their first restriction during the previous three months, separating neighboring counties from those that share no border. Separate regressions for these two connections compare counties within the same state and month and account for the county's own announcement status, unresolved project dates and the number of neighboring counties that adopted earlier. The model for counties that share no border also includes a connection based on assignment to the same news outlet (SI S6; Table S9).

For data center construction, we compared projects in counties that adopted a restriction within six months of announcement with other projects whose announcement month is known. The sample contains 898 projects with a county assignment and an announcement month: 388 were recorded as started by October 2026, 415 as announced and 95 as cancelled. We define a project as exposed when its county adopted a restriction during the announcement month or the five months after it. 110 projects meet this definition, 86 of which were in 2026. We regress recorded construction status on this county exposure, announcement-month effects, the logarithm of one plus reported megawatts, and an indicator for missing capacity. The model compares projects announced in the same month and accounts for differences in reported size; standard errors come from 999 county-level bootstrap draws (Table 4).

A second regression divides the exposed projects according to whether an opposition action was recorded in the county before its first restriction, using the Data Center Opposition

Tracker, movement records, the Crowd Counting Consortium and ACLED, and directly estimates the difference between those two groups. SI Tables S6 to S8 report alternative three-, nine- and twelve-month exposure windows, shifted policy dates, equal follow-up, bounds on twelve-month construction shares, and monthly models of starts and cancellations. The replication package lists the 175 pairings of 110 projects with 134 distinct instruments.

## 4. Findings

### *4.1 Geography of restrictions*

Our first finding is that most counties that adopted a restriction did so without a new data center project being imminent. Of the 445 adopting counties, 256 fall in this group, while 147 had a project announced in or before the month of their first restriction (another 33 had projects with unclear dates). Counties with a recorded project adopted restrictions at nearly three times the rate of counties with no recorded project: 189 of 703 (c. 27%), compared with 256 of 2,519 (c. 10%). Of the 189 counties with both a data center project and a restriction, 147 already had a project on record by the month they adopted their first restriction, in 9, the first recorded project came later (indeterminate for 33). The majority of adopting counties had no recorded project, supporting H2. The restrictions are concentrated in 2026. Of the 603 in the main series, 534 carry a 2026 adoption date, 58 a 2025 date and 11 an earlier date, classifying 502 restrictions as active, 46 as extended, 30 as replaced, 19 as expired and six as rescinded.

*Table 2. Project history at first recorded policy adoption*

| **Project history** | **Counties** | **Percent** |
|---|---|---|
| No project recorded | 256 | 57.5 |
| Known project at or before adoption | 147 | 33.0 |
| Project timing unresolved | 33 | 7.4 |
| Projects recorded only after adoption | 9 | 2.0 |
| Total | 445 | 100.0 |

*4.2 Timing of first restrictions after announcement*

Counties most often adopted their first restriction in the announcement month or the five months after it. 33 first restrictions fell in that period, nine in the next six months and twelve in months twelve to twenty-four. Comparing counties within the same calendar month, the estimated increase in monthly adoption after announcement is 0.89 percentage points, with a 95 percent interval of 0.53 to 1.25. See supplement for alternative specifications. Most of the increase occurs in the first six months: 1.76 points in months zero to five, compared with 0.38 in months six to eleven and 0.49 from month twelve onward (Figure 1), with both 95 percent intervals for months 0-5 and beyond month 12 above zero. First adoption is otherwise rare across the panel. There were 325 such decisions across 326,374 county-months, about one adoption for every 1,000 monthly observations. See Table 3 and supplement S11+S16.

## Table 3. First recorded policy adoption

| **Dependent variable: first adoption** | **(1) Month** | **(2) State-month** | **(3) County + month** | **(4) Timing bins** |
|---|---|---|---|---|
| Post announcement | 0.89<br>(0.18) | 0.81<br>(0.18) | 1.09<br>(0.22) | |
| Months 0-5 | | | | 1.76<br>(0.39) |
| Months 6-11 | | | | 0.38<br>(0.30) |
| Month 12 onward | | | | 0.49<br>(0.23) |
| Calendar-month effects | Yes | No | Yes | Yes |
| State-by-month effects | No | Yes | No | No |
| County effects | No | No | Yes | No |
| County-months | 326,374 | 326,374 | 326,374 | 326,374 |
| County clusters | 2,781 | 2,781 | 2,781 | 2,781 |
| First adoptions | 325 | 325 | 325 | 325 |

Note: OLS coefficients in percentage points per county-month; county-clustered standard errors in parentheses. Each column is a separate regression. Column 4 includes the three timing indicators together, comparing those periods with pre-announcement observations and counties with no recorded project. Absorbed effects include the intercept. SI S3 and S11 report 95% intervals, alternative inference and direct comparisons between periods.

**Figure 1. Timing of recorded policy adoption**

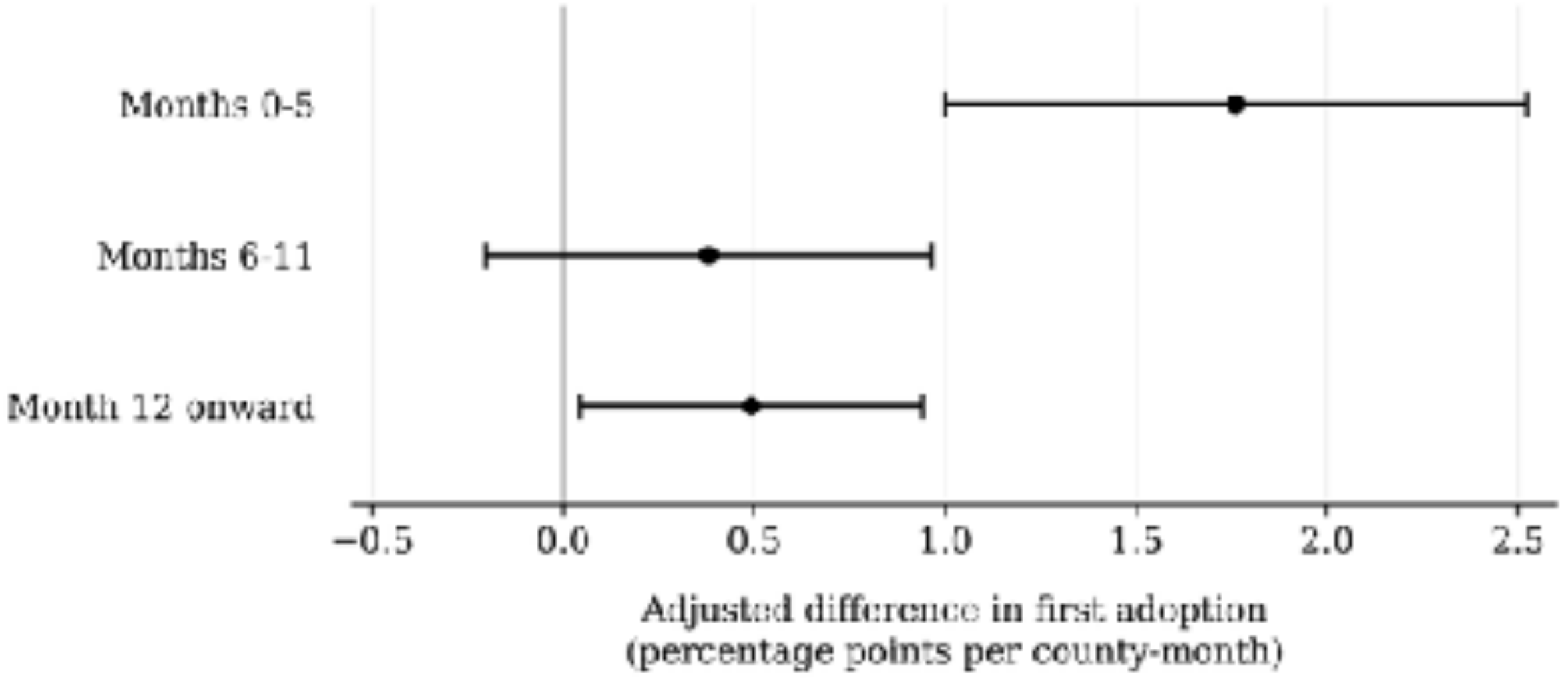

Local reporting on data center contention sometimes, but not often, preceded the first recorded project announcement. Of the 274 announcement counties, 236 have an outlet website assigned to that county in the 3DLNews directory. In 43 of these counties, we found contention reporting during the twenty-four months before announcement; the median lead was ten months (SI S10). Articles enter this count through the outlet's county assignment in the directory, so a report can concern a dispute elsewhere. Nine counties also adopted their first restriction before the announcement month, five of them in the preceding two months (SI Table S16). Residents also had a better chance of finding out about a proposed data center from the news in counties with more local outlets. By the month a project first appeared in our records, local or regional journalists had reported on it in 60% of counties with at least three listed outlets, compared with 40% of counties with fewer. In 15 counties, reporters gave the date of an upcoming hearing, vote or public meeting, with a median of 12 days' notice. Residents attended, asked questions or spoke in 11 of those counties. The greater opportunity to learn about a proposal through local reporting gives H5 indirect support.

*4.3 Adoption among counties sharing opposition groups*

There is a snowball effect to restrictions. Counties with similar opposition group as counties where restrictions have been enacted are more likely to adopt restrictions themselves. Across the panel, there is about one first adoption for every 900 county-months. The rate is much higher during the three months after a connected county adopts. Five of 54 county-months following adoption by a neighboring county with a shared group ended in a first restriction. For connected counties that shared no border, fourteen of 202 county-months ended in adoption (SI Table S10). The regression estimates how monthly adoption differs. For a neighboring county with a shared group, the estimate is 5.76 percentage points, with a 95 percent interval of -1.22 to 12.73. For a connected county that shares no border, it is 1.41 points, with an interval of -0.15 to 2.97. These comparisons account for state and month, the county's own project status and earlier adoptions by its neighbors, using the controls in Section 3.2 (SI Table S9). Fourteen of the sixteen distinct adoption events occurred where the organizational connection crossed a gap between counties, including three events also connected to a neighboring adopter (SI Table S10). The estimated increase for neighboring counties with a shared group is in the direction H3 predicts, but the test is inconclusive.

Counties with and without documented coverage adopted restrictions at similar rates in the five months following the announcement month (11 of 82 with, 10 of 82 without), among counties that had yet to adopt and had full follow-up. By the announcement month, citizen action had already been documented in 22 of 95 counties with verified media coverage and 9 of 96 without it. Among counties sharing a website with a recent adopter that shared no border, nine of 280 county-months ended in adoption. The corresponding count for the two organizational connections combined is sixteen adoptions across 230 county-months. Accounting for the other variables in the regression gives an estimated difference of 0.88 percentage points for the shared-website connection, with a 95 percent interval of -0.90 to 2.67 (SI Tables S9 and S10). We also examined nine pairs of counties that shared a news outlet and adopted restrictions in succession. Iowa Public Radio, for example, reported in July 2026 that Linn County had paused new data center applications. Seven weeks later, the Waterloo City Council in Black Hawk County adopted a pause of its own. Residents in Waterloo had pressed for the measure after learning that their city had already offered a developer tax breaks and land for one dollar.

*4.4 Construction status and early county restrictions*

Fewer projects had started construction in counties that adopted a restriction within six months of announcement. By October 2026, eighteen of the 110 projects in this group had a recorded start, compared with 370 of the other 788 projects, or sixteen percent against 47 percent. The two groups differ in how long they had to begin construction: 86 of the 110 exposed projects were announced in 2026, giving them at most ten months of follow-up. After accounting for announcement month and reported capacity, projects in counties with early restrictions were 7.89 percentage points less likely to have started (95 percent interval for the difference: -16.75 to 1.99; Table 4, column 1). The difference remains negative with alternative exposure windows, shifted policy dates and equal follow-up, with all intervals spanning zero (SI Table S6).

The 110 exposed projects divide into 71 in counties with an opposition action recorded before their first restriction and 39 in counties with no prior action recorded in the four sources. Seventeen of the 71 projects in the first group had started construction, compared with one of the 39 in the second (SI S10). After accounting for announcement month and capacity, projects in counties with a prior recorded action were 1.35 percentage points less likely to have started than the other 788 projects. For projects in counties with no prior recorded action, the difference was 20.16 points. The estimated difference between the two exposed groups is 18.81 points, with a 95 percent interval of 6.67 to 31.12 (Table 4, column 2). H4 is therefore not supported as stated. A smaller share of projects started construction in counties with early restrictions, but the largest gap was in counties with no recorded opposition before the vote.

**Table 4. Regressions of recorded construction status**

| **Dependent variable: construction started** | **(1)<br>Pooled exposure** | **(2)<br>Prior-opposition split** |
|---|---|---|
| Early county restriction | -7.89<br>(4.57) | |
| Early restriction, prior action recorded | | -1.35<br>(5.53) |
| Early restriction, no prior action recorded | | -20.16<br>(4.84) |
| Log(1 + reported MW) | -0.33<br>(1.19) | -0.21<br>(1.19) |
| Capacity missing | -14.87<br>(7.25) | -13.89<br>(7.27) |
| Announcement-month effects | Yes | Yes |
| Projects | 898 | 898 |
| County clusters | 513 | 513 |
| Recorded construction starts | 388 | 388 |
| Exposed projects | 110 | 71 / 39 |

Note: OLS coefficients in percentage points; county-bootstrap standard errors in parentheses (999 valid draws; seed 20260906). The pooled 95% interval is [-16.75, 1.99]. SI S4 and S10 report sensitivity and chronology checks.

Twenty-four of the 110 exposed projects were classified as cancelled by October 2026, fourteen in counties with prior recorded opposition and ten in the other group. Comparing each assigned cancellation month with the first restriction in that project's six-month exposure window, the restriction came first in eight cases, both fell in the same month in six, and cancellation came first in ten. Nine of the cancellation dates come from the report documenting the cancellation, which places the event at or before that report's month. SI Table S8 reports monthly models of starts and cancellations using the dated events. SI Table S7 gives bounds on twelve-month construction shares for the 487 projects with a full year of follow-up through September 2026; fifteen are exposed, and 162 construction starts in the comparison group and three in the exposed group have no recorded month.

*4.5 Direct voter decisions on data center development*

Local resistance prevailed in every data-center-specific question identified on the November 3, 2026 election. Across 42 questions in 39 jurisdictions and eight states, every reported outcome favored rejection or restriction. Support for the restrictive position ranged from 60.1% to 88.3% of valid votes, with a median of 77.2%. Four separate questions appeared on the same ballot in Wilmington, Ohio, so the collection contains 42 decisions by 39 electorates (32 binding and 10 advisory). Every explicit invitation to endorse development failed. Binding defeats covered everything from Boulder City's land-use permission question, Ashville's resolution on development negotiations, and four separate Wilmington rezonings. Voters also approved all 26 binding questions proposing prohibitions, moratoria or developer restrictions, together with three advisory questions supporting restrictions (see Table 5).

This resistance crossed partisan lines. Using the 2024 presidential vote to identify local partisan lean, 29 of the 38 matched jurisdictions voted Republican (76.3%) and nine Democratic (23.7%). These places accounted for 41 ballot questions, split 32 (78.0%) in Republican-leaning jurisdictions and nine (22.0%) in Democratic-leaning jurisdictions; every outcome favored rejection or restriction. For example, in Trenton, OH, where Trump received 77.0% of the two-party presidential vote, 79.7% supported the restrictive position; in Granville, OH, where Harris received 64.7%, the corresponding share was 82.4% (MIT Election Data and Science Lab 2026; authors' compilation). The specific initiatives varied, from advisory votes to express preferences, votes to consider existing laws, social and environmental impacts, and economic / tax implications. Regardless of specifics, the resistance was absolute.

**Table 5. Pro- and anti-data center outcomes, November 3, 2026**

| Question type | Questions | Pro outcomes | Restrictive outcomes |
|---|---|---|---|
| Binding prohibitions, moratoria or developer conditions | 26 | 0 | 26 |
| Binding development permissions or project arrangements | 6 | 0 | 6 |
| Advisory support for restrictions | 3 | 0 | 3 |
| Advisory support for development | 7 | 0 | 7 |
| **Total** | **42** | **0** | **42** |

Note: Counts refer to ballot questions. Anti-data center outcomes comprise approval of restrictions or conditions and rejection of development permissions or support; the reverse is coded pro-data center.

## 5. Discussion

To contexualize our findings, we return to foundational studies. Molotch (1976) explains why local landowners and officials join forces to attract investment, while Burrell (2020) shows how early that work can begin. The findings support the application of this work for the data center era. Just like how municipal leaders in Prineville, Oregon, set aside land for economic development decades before Facebook stepped in, communities can use the same planning powers to limit what investments they want to accept. A council that restricts data centers with no recorded project preemptively sets terms that data center developers will see as evidence that the community is aware of the issues at play. But if a firm names a site before such restrictions, the political calculus is different; in these cases, citizens are playing catch-up to a decision the firm has already taken, and officials must answer for what they are prepared to grant a particular applicant. The concentration of first restrictions soon after announcements therefore makes sense alongside the many restrictions adopted in places with no project. In both settings, the council exercises power over mobile capital, including through decisions made before there is a firm at the bargaining table.

A firm's threat to build elsewhere puts pressure on officials who want its investment; officials seeking to keep a facility out achieve their objective when it leaves. Local restrictions also differ in which projects they cover. For example, Birmingham paused new applications for facilities of 20 megawatts or more filed after March 3, 2026, while preserving the rules for two projects already in permitting. Independence's pause covers new data center and battery storage applications and exempts the Nebius project already under review (City of Birmingham 2026; City of Independence 2026). These councils retained the ability to review rules for future applications while allowing the named projects to continue through the existing process. In the national construction comparison, the largest difference is among the 39 projects in counties where the restriction preceded any opposition action recorded in our sources, with one project recorded as started by October 2026. State legislatures also decide how much authority councils retain over such development. The state-policy file contains 438 data center bills, of which 399 have a last action in 2025 or 2026, placing the local decisions within the wider conflict over state and municipal authority (Goodman et al. 2021; SoRelle and Fullerton 2024; Kollar 2026b).

The most surprising finding was related to eventual groundbreaking. Among data center projects facing an early restriction, we expected construction starts to be lower where residents had organized. In fact, even where residents had organized and officials had adopted a restriction, projects broke ground more often than we expected

The assumption was that residents would organize after hearing about a new data center to stop it, although research on the US pension movement suggests that assessing a campaign's influence requires knowing which decision its members wanted to change and which officials had the power to change it (Amenta et al., 2005). In Birmingham, for example, the council's moratorium on new applications for large data centers gave residents time to press for a say in where future facilities could be built. For anyone seeking to stop the DC Blox expansion or the Nebius AI Factory, the same vote left that demand unresolved because both projects could proceed under existing rules (City of Birmingham, 2026). This finding may reflect the fact that the filing of a permit application gave residents something concrete to organize against, while also putting existing/planned projects beyond the reach of a moratorium that covered only new applications.

Beginning with a disputed facility determines which local decisions enter a study of siting politics. Research has used those disputes to examine who opposed development and what happened to the proposal (Giordono et al. 2018; Stokes et al. 2023; Reece et al. 2026). Our project inventory covers 189 of the 445 adopting counties; the policy records identify another 256 with no recorded project. Studying decisions in those counties requires starting with the local government, using policy documents and meeting records such as those collected by LocalView (Barari and Simko 2023). The same records offer a way to trace what officials learned through organizations working across counties, including how an earlier dispute shaped their understanding of the problem and their choice of response (Gilardi 2010; Gilardi et al. 2021). Our organizational dates establish when a group had been recorded in both counties, while meeting accounts and interviews provide the detail of what participants exchanged. Local reporting supplies another dated account of contention, including the coverage preceding project announcements in 43 counties, and connects public decisions to the electoral accountability studied in research on the local press (Snyder and Strömberg 2010; SI S10).

Culpepper (2010) argues that firms exercise greater influence when political decisions attract little public attention. In data center settings this also holds. Before a vote, a report on a tax agreement lets residents question what their council is about to offer. Afterward, the same information allows them to hold officials to account for a commitment already made. This helps explain why the association between more local outlets and early coverage matters even though councils with and without media coverage subsequently adopted restrictions at similar rates. As noted previously, Arnold et al. (2017) examine how advocates' relationships shape municipal decisions on fracking, and homeowners with time to attend hearings and file comments are overrepresented in land-use decisions and use those procedures to oppose nearby development (Einstein et al. 2019). Applied here, their argument turns on whether organizers used one county's experience to persuade another council to act. Early reporting identifies communities where residents had a chance to take part, whichever policy they favored.

Therefore, the larger question of "can citizens stop a data center?" can be answered in the qualified affirmative. A given citizenry is more likely to be successful when it preemptively passes legislation against data centers, when the local media environment is rich and responsive, and when neighboring communities have also been successful. It is also more likely to be successful if it acts quickly (within the first six months of the notification), but after permitting is issued the influence discussion shifts to the terms of the deal, not its existence. This research has forward value for research on various aspects of business-government-citizen relationships, and on the role of technology actors in society.

### *5.1 Limitations*

Announcements, opposition and restrictions respond to the same development pressures, and problems with financing or grid connections affect a project's prospects before a council acts. The comparisons measure associations within those sequences. Ten exposed projects have cancellation dates before the relevant restriction and six have cancellation and restriction in the same month. Month-level dates leave the order of a hearing, application and vote within a month unresolved, while developers choose when to announce publicly and sometimes do so after residents and officials have learned of the proposal. Construction dates are also incomplete: of eighteen exposed projects recorded as started, fifteen have no resolved start month, two began before the relevant restriction and one began after it. Eighty-six of the 110 exposed projects were announced in 2026, leaving a short period in which to observe construction by October. The pooled construction estimate has a 95 percent interval from -16.75 to 1.99 percentage points. Longer follow-up is needed to distinguish projects that eventually proceed from those abandoned or moved; the twelve-month comparison currently contains fifteen exposed projects among 487 with a full year of observation (SI Tables S6 and S7).

Documentation is uneven across places and years: the policy inventory's systematic search ran from May to July 2026, and its authors flagged 234 of the 603 eligible records for source verification. Language-model coders checked source excerpts for a sample of 100 tracker records and found full support for the named group in seven and partial support in ten (SI Table

S13). AI answer engines studying armed conflicts with sparse reporting made more factual errors than they did on extensively covered conflicts (Miklian 2026a). A group's date marks its first appearance in the sources for a county; establishing when the group began its work requires a separate history of the campaign. Groups' choices about where to work also shape these connections, since an organization has reasons to concentrate its efforts in receptive counties. The adjacent and nonadjacent estimates rest on five and fourteen adoption events, respectively, and both county-clustered intervals include zero. Establishing what the second council learned from the first requires the documents or interviews recording that exchange. The four opposition sources differ in collection procedures and geographic coverage, making the split in Table 4 depend on which actions they captured before adoption, as research on protest data illustrates (Dorff et al. 2023; Fisher et al. 2019; Wang et al. 2016). Publication dates show when readers could learn of a proposal, but only individual accounts establish when a resident first heard of it.

The construction measure matches projects and restrictions within a county, whereas legal coverage depends on the authority adopting the rule, the land it governs, the regulated use, operative dates and application history. The review file contains 175 pairings of 110 projects with 134 distinct instruments; the pairings involve 80 city, 39 county, 31 township, fourteen town, nine village and two other entries (SI S8). A city ordinance applies within the city's territory, leaving projects elsewhere in the county under their own jurisdiction's rules. Checking the dates also establishes whether a restriction had expired or been replaced before the project decision. Independence expressly exempts the Nebius project already under review, yet the project meets our county exposure definition. Removing that project changes the estimated construction difference to -8.76 percentage points, with a 95 percent interval of -17.69 to 0.96 (SI

Table S6). The institutional argument rests on American land-use arrangements, in which counties and municipalities exercise local authority and residents contest applications through hearings. Authority is allocated differently where a national agency selects sites, and it also varies within our sample of 52 state-level jurisdictions.

**AI Use Disclosure Statement:**

The authors used Claude Opus 4.8 (Anthropic) as an adversarial reviewer to assess a first draft of this document, to build tables and verify data, and to take a final copyediting / references pass. OpenAI Codex Astra was used to verify and audit the replication package against the underlying data, e.g. in coding and denominator corrections and ensuring consistency between the paper and replication materials. All AI-assisted output was carefully reviewed and accepted by the authors, who take full responsibility for all material herein. Neither tool was used to develop the primary argumentation or to assist in the authors' substantive interpretation.